\documentclass[twocolumn,english,superscriptaddress,pre]{revtex4-1}
\usepackage{xcolor}
\usepackage{bm}
\usepackage{amstext}
\usepackage{amssymb}
\usepackage{graphicx}
\usepackage{commath}
\usepackage{dsfont}
\usepackage{mathtools}
\usepackage{amsmath,mathrsfs}
\usepackage{physics}
\usepackage{hyperref}
\hypersetup{colorlinks=true, linkcolor=blue, citecolor=blue, urlcolor=blue, unicode=true}

\begin{document}

\title{U(1) symmetric recurrent neural networks for quantum state reconstruction}

\author{Stewart Morawetz}
\affiliation{Department of Physics and Astronomy, University of Waterloo, Ontario, N2L 3G1, Canada}

\author{Isaac J.S. De Vlugt}
\affiliation{Department of Physics and Astronomy, University of Waterloo, Ontario, N2L 3G1, Canada}
\affiliation{Perimeter Institute for Theoretical Physics, Waterloo, ON N2L 2Y5, Canada}

\author{Juan Carrasquilla}
\affiliation{Vector Institute, MaRS Centre, Toronto, Ontario, M5G 1M1, Canada}
\affiliation{Department of Physics and Astronomy, University of Waterloo, Ontario, N2L 3G1, Canada}

\author{Roger G. Melko}
\affiliation{Department of Physics and Astronomy, University of Waterloo, Ontario, N2L 3G1, Canada}
\affiliation{Perimeter Institute for Theoretical Physics, Waterloo, ON N2L 2Y5, Canada}

\date{\today}

\begin{abstract}

Generative models are a promising technology for the enhancement of quantum simulators.
These machine learning methods are capable of reconstructing a quantum state from experimental measurements, 
and can aid in the calculation of physical observables.
In this paper, we employ a recurrent neural network (RNN) to reconstruct the ground state of the spin-1/2 XY model,
a prototypical Hamiltonian explored in trapped ion simulators.
We explore its performance after enforcing a U(1) symmetry, which was recently shown by 
Hibat-Allah {\it et al.} \cite{Hibat-Allah2020} to preserve the autoregressive nature of the RNN.
By studying the reconstruction of the XY model ground state from projective measurement data,
we show that imposing U(1) symmetry on the RNN significantly increases the efficiency of learning, particularly 
in the early epoch regime.
We argue that this performance increase may result from the tendency of the enforced symmetry to 
alleviate vanishing and exploding gradients, which helps stabilize the training process.
Thus, symmetry-enforced RNNs may be particularly useful for applications of quantum simulators where
 a rapid feedback between optimization and circuit preparation is necessary, such as in
hybrid classical-quantum algorithms.

\end{abstract}

\maketitle

\section{Introduction}

Advances in the fabrication and control of quantum devices is progressing to the point where high-quality measurement
data is attainable from extremely pure quantum states of tens or hundreds of qubits.  A growing body of work has demonstrated
the ability of technologies adopted from machine learning to provide approximate reconstructions of quantum states from practically-sized 
data sets on such noisy intermediate-scale quantum (NISQ) devices
\cite{torlaiLearningThermodynamicsBoltzmann2016,torlaiNeuralNetworkQuantumState2018,RevModPhysML,Juan_review}. 
A typical machine learning approach involves the use of generative models to learn and represent a multi-qubit quantum state. 
The most well-studied generative model in this context is the restricted Boltzmann machine (RBM), which has been used to obtain 
high-quality reconstructions of a variety of pure and mixed states from both synthetic and experimental data \cite{LukinPRL,Melko_NatPhysPersp,ARCMP}.

Despite their success, one disadvantage of RBMs well-known to the machine learning community is that they do not provide a tractable distribution estimator -- in other words, a normalized representation of the quantum state.
This has motivated physicists to explore powerful alternatives in the so-called autoregressive models \cite{carrasquillaReconstructingQuantumStates2018,Dian_Wang_Zhang,Variational_autoregressive,Roth}, 
such as recurrent neural networks (RNNs) or attention-based transformers \cite{AQT}. Autogregressive models, where the total (joint) distribution of the variables is decomposed into a product of conditional probabilities, are capable of learning a normalized representation of the quantum state. Additionally, they provide a tractable method capable of efficiently producing perfectly independent samples, e.g.~for use in calculating estimators of physical observables. 

Beyond normalization, another important aspect of improving the performance of any quantum algorithm is the exploitation of quantum symmetries.  More specifically, Hamiltonians that are invariant under a symmetry group will have conserved quantities which can be exploited to significantly simplify the task of computing the ground state wavefunction. A well-known example of a performance increase comes from the implementation of the U(1) symmetry of particle number (charge) or magnetization sector.  We will focus on a simple model Hamiltonian invariant under U(1) symmetry operations in this paper, the spin-1/2 XY model, which has recently been engineered in trapped ion experiments \cite{Ions1,Ions2,Ion_Review}.
The Hamiltonian of the model is,
\begin{equation} \label{eq:xy_hamiltonian}
    H =  -J \sum_{\langle i j \rangle} (S_i^x S_j^x + S_i^y S_j^y), 
\end{equation}
where $\langle i j \rangle$ denotes nearest-neighbor pairs of a one-dimensional lattice with $N$ sites, and ${\bf S}_i$ is the spin-1/2 operator on site $i$.
The U(1) symmetry of the Hamiltonian is manifest as a block diagonal structure, and a conservation law for the total magnetization $S^z=\sum_{i} S^{z}_{i}$ of the ground state.

As recognized by M. Hibat-Allah {\it et al.}~\cite{Hibat-Allah2020}, it is possible to implement a U(1) symmetry on an RNN wavefunction while preserving its autoregressive property, through a projection operator on the conditional probability distributions that are the output of the RNN cell (described in detail below).  In Ref.~\cite{Hibat-Allah2020}, the symmetry-conservation projection was employed in an RNN used as a variational wavefunction. In that context, the RNN cost function is defined by the expectation value of the Hamiltonian -- a calculation which is improved significantly by the implementation of U(1) symmetry.  In the present work, we explore the implementation of this symmetry in an RNN used for data-driven state reconstruction.  In order to do so, we generate training and testing data from the Hamiltonian Eq.~\eqref{eq:xy_hamiltonian}, and train the parameters of an RNN to optimize a negative log-likelihood.  
By comparing and contrasting to both RBMs and RNNs without any implementation of symmetry, we present evidence that RNNs constrained to obey the U(1) symmetry
result in a significant speed up in the time required to train to low infidelity and relative energy difference,
particularly in the early epoch training regime.

\section{Recurrent Neural Networks} \label{sec:RNN_info}

The RNN is a powerful autoregressive model built for modelling sequential data,
which has been discussed in many other works; we will leave the reader to peruse references relevant for its
application to machine learning problems in industry applications \cite{lipton2015critical}.
The RNN was introduced into the field of quantum state reconstruction in Ref.~\cite{carrasquillaReconstructingQuantumStates2018}, 
and as an explicit wavefunction representation for variational optimization in Ref.~\cite{Hibat-Allah2020}.
In this section, we offer a brief background of RNNs for quantum state reconstruction as used in the present study.

\subsection{RNN architecture} \label{sec:RNN_architecture}

The goal of an RNN is to reconstruct an unknown target probability distribution $q(\bm{\sigma})$ by interpreting instances of an input data set $\mathcal{D} = \{\bm{\sigma}\}$, sampled from $q(\bm{\sigma})$ as sequential data.
In this paper, $\bm{\sigma}$ is a vector of size $N$ containing the occupation number of each site $i$, which we relate to the eigenvalues of the 
spin operator $S_i^z = \pm 1/2$ through $\sigma_i = 1/2 - S_i^z$.

In the context of quantum state reconstruction, the target distribution is the square of the ground state (GS) of Eq.~\eqref{eq:xy_hamiltonian},
\begin{equation}
    q(\bm{\sigma}) = \vert \braket{\bm{\sigma}}{\psi_{\text{GS}}} \vert^2 = \vert \psi_{\text{GS}}(\bm{\sigma})\vert^2.
\end{equation}
Setting $J=1$ in Eq.~\eqref{eq:xy_hamiltonian}, the Hamiltonian satisfies the well-known Perron-Frobenius theorem \cite{tarazagaPerronFrobeniusTheorem2001}.  Therefore, the ground state is sign-free and can be written as the square root of the target distribution,
\begin{equation} \label{eq:exact_groundstate}
    \psi_{\text{GS}}(\bm{\sigma}) = \sqrt{q(\bm{\sigma})}.
\end{equation}
Since the wavefunction has no complex phase, it can be reconstructed from projective measurements drawn from the $S^z$ basis only. 
For use in the RNN, each $\bm{\sigma}$ is formatted as a one-hotted configuration %in the $S^z$ basis from a many-body Hilbert space 
comprising of $N$ ``sequential'' qubits. Specifically, $\bm{\sigma} \equiv (\sigma_1, \sigma_2, ... , \sigma_N)$ with $\sigma_i \in \{0,1\}$ or $\bm{\sigma}_i \in \{(1,0), (0,1)\}$ in the binary and one-hotted representations, respectively. Throughout this work, we adopt the convention that spin-up refers to $\sigma_i = 0$ and spin-down refers to $\sigma_i = 1$.

The RNN defines an autoregressive model for $p(\bm{\sigma})$ using the chain rule of probabilities,
\begin{equation} \label{eq:RNN_probability}
    p(\bm{\sigma}) = p(\sigma_1) p(\sigma_2 | \sigma_1) ... p(\sigma_N | \sigma_{N-1}, ..., {\sigma}_1).
\end{equation}
The building block of an RNN is a recurrent cell \cite{lipton2015critical}. A ``vanilla'' RNN consists of a nonlinear activation function $f$ that maps a hidden vector $\bm{h}_{i-1}$ of dimension $d_h$ and an input vector $\bm{\sigma}_{i-1}$ of dimension $d_v = 2$ to a different hidden vector $\bm{h}_i$. Specifically, 
\begin{equation} \label{eq:rnn_layer}
    \bm{h}_i = f\left(W\bm{\sigma}_{i-1} + U\bm{h}_{i-1} + \bm{b}\right),
\end{equation}
where tunable parameters $\bm{\theta}$ of the RNN are given by the weight matrices $W \in \mathbb{R}^{d_h \times 2}$, $U \in \mathbb{R}^{d_h \times d_h}$, and bias $\bm{b} \in \mathbb{R}^{d_h}$.
The initial values of the hidden and input vectors $\bm{h}_0$ and $\bm{\sigma}_0$ are fixed to the zero vector and $(1,0)$, respectively.

\begin{figure}  % Standard RNN structure
    \centering
    \includegraphics[width=\linewidth]{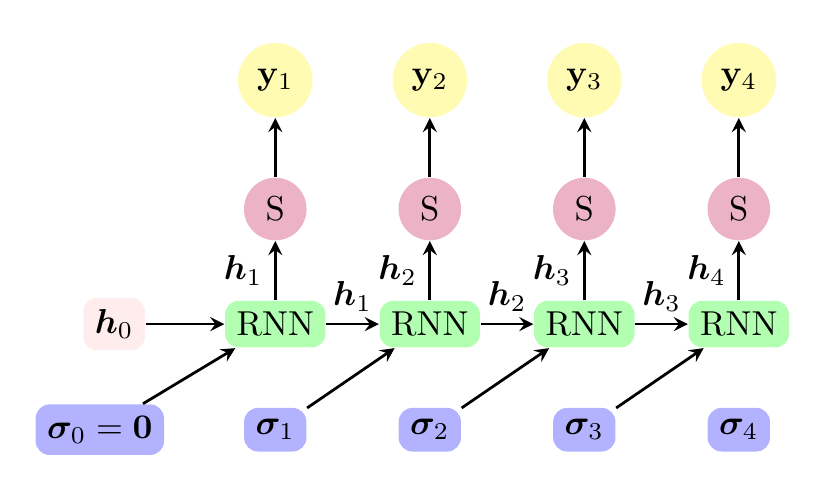}
    \caption{A diagram of the RNN architecture. At the $i$-th recurrent step, the RNN cell takes the previous hidden and input vectors $\bm{h}_{i-1}$ and $\bm{\sigma}_{i-1}$, and outputs the next hidden vector $\bm{h}_i$. This is then passed to a fully connected softmax layer $S$, which makes a prediction $\bm{y}_i = (p(\sigma_i = 0), p(\sigma_i = 1))$ for the next input $\bm{\sigma}_i$. $\bm{h}_i$ and $\bm{\sigma}_i$ are passed to the next RNN recurrent cell in order to compute $\bm{h}_{i+1}$. This is repeated until there are output vectors $\bm{y}_1$, $\bm{y}_2$, ..., and $\bm{y}_N$ corresponding to probabilities $p(\sigma_1)$, $p(\sigma_2 | \sigma_1)$, ..., and $p(\sigma_N | \sigma_{N-1}, ..., \sigma_1)$, respectively.}
    \label{fig:standard_RNN}
\end{figure}

A linear layer then maps each output hidden vector $\bm{h}_i$ to an output vector $\bm{y}_i \in \mathbb{R}^2$, where 
\begin{equation} \label{eq:softmax_layer}
    \bm{y}_i = S\left(V\bm{h}_i + \bm{c}\right),
\end{equation}
$S$ is the softmax function,
\begin{equation*} \label{eq:softmax_def}
    S(v_j) = \frac{\exp(v_j)}{\sum_i \exp(v_i)},
\end{equation*}
and additional tunable RNN parameters are given by the weight matrix $V \in \mathbb{R}^{d_h \times 2}$ and bias $\bm{c} \in \mathbb{R}^2$. This entire structure is shown diagramatically for $N = 4$ in Fig.~\ref{fig:standard_RNN}.

This  vanilla RNN cell architecture has been shown not to lend itself well to learning longer sequences of data, as gradients tend to explode or vanish \cite{Bengio1994}. To mitigate this, we use the slightly modified gated recurrent unit (GRU) cell, which has demonstrated stronger performance with learning long sequences of data \cite{cho-etal-2014-properties, carrasquillaReconstructingQuantumStates2018,Hibat-Allah2020}. Additional details of this architecture can be found in Appendix \ref{appendix:GRUs}. 

For the purposes of quantum state reconstruction, 
the output $\bm{y_i}$ in Eq.~\eqref{eq:softmax_layer} is interpreted as a probability vector corresponding to the $i$-th qubit having $\sigma_i = $ 0 or 1 given all of the previous inputs $\sigma_{<i}= \sigma_{i-1}, \sigma_{i-2},\ldots \sigma_1$,

\begin{equation*} \label{eq:intepret_output}
    \bm{y}_i = \big(p(\sigma_i = 0 | \sigma_{<i}), \hspace{1mm} p(\sigma_i = 1 | \sigma_{<i}) \big).
\end{equation*}
Then, Eq.~\eqref{eq:RNN_probability} is calculated as 
\begin{equation}
    p(\bm{\sigma}) = \prod_{i=1}^N \bm{y}_i \cdot \bm{\sigma}_i,
\end{equation}
and the RNN wavefunction is given by
\begin{equation} \label{eq:RNN_wavefunction}
    \psi_{\text{RNN}}(\bm{\sigma}) = \sqrt{p(\bm{\sigma})}.
\end{equation}
The goal of the RNN is to find a $p(\bm{\sigma})$ which approximates the unknown target distribution $q(\bm{\sigma})$
as accurately as possible, given only the data set $\mathcal{D}$.
In order to achieve this, the RNN is trained by tuning the parameters $\bm{\theta}$ (the weights and biases)
in such a way as to minimize the negative log-likelihood (NLL), which is given by
\begin{equation} \label{eq:NLL}
    \text{NLL} = -\frac{1}{|\mathcal{D}|} \sum_{\bm{\sigma} \in \mathcal{D}} \log p(\bm{\sigma}).
\end{equation}
This NLL thereby defines the loss landscape of the RNN.
For all RNNs trained in this work, stochastic gradient decent (SGD) is employed to perform the optimization numerically.

\subsection{Sampling from RNNs and symmetry enforcement}

\begin{figure} 
    \centering
    \includegraphics[width=\linewidth]{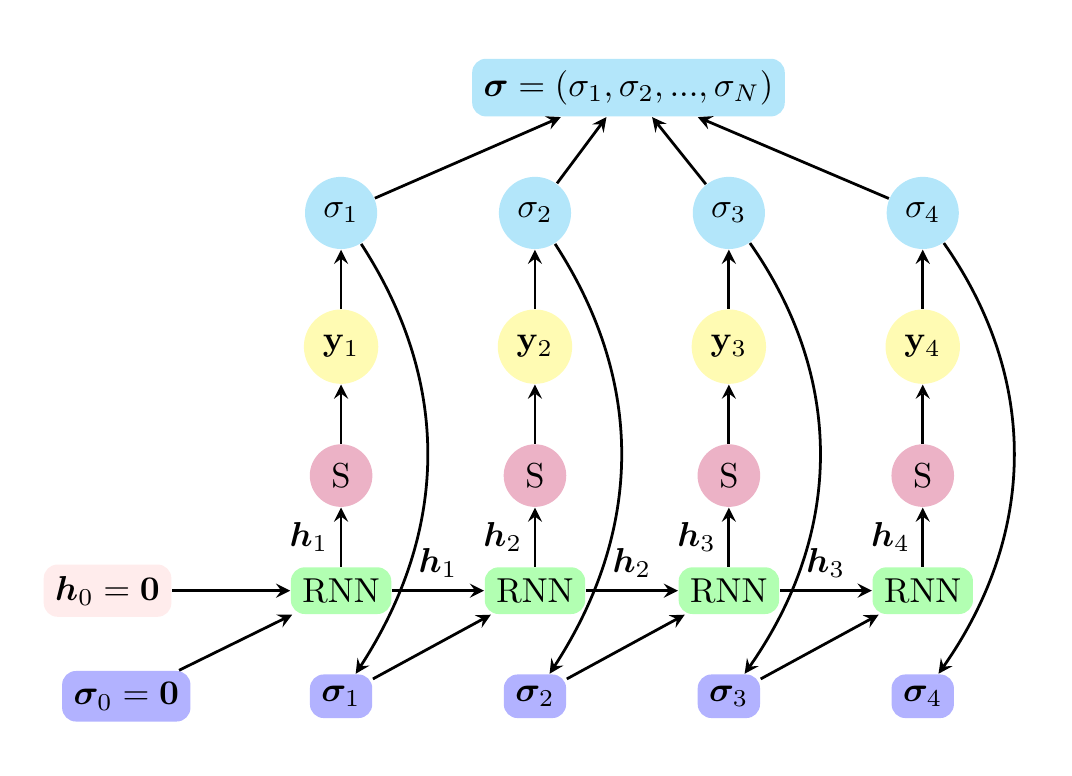}
    \caption{A diagramatic depiction of sampling from an RNN for $N=4$ spins. 
    At each recurrent step, the RNN uses the output $\bm{y}_i$ as a probability distribution from which to sample the next input spin, $\sigma_i$. This is then one-hotted and fed into the next RNN recurrent cell to calculate the proceeding spin value $\sigma_{i+1}$, and so on. In this way, the RNN can generate uncorrelated samples $\bm{\sigma}$ autoregressively.}
    \label{fig:autoreg_RNN}
\end{figure}

In order to discuss the implementation of symmetry in RNN training, it is instructive
to first consider how one draws samples from the RNN probability distribution $p(\bm{\sigma})$. 
Figure \ref{fig:autoreg_RNN} illustrates the general process of sampling from an RNN.
At each recurrent step, the output probability distribution $\bm{y}_i$ is sampled to obtain the input $\sigma_i$ to the proceeding RNN recurrent unit. 
In this way of autoregressive sampling, the RNN can generate perfectly uncorrelated samples of $p(\bm{\sigma})$ in a straightforward and efficient manner.
While this efficiency is a main advantage of this generative model architecture, it is clear that the $N$ samples of $\sigma_i$ thus produced obey 
no global constraints, such as the conservation of $S^z$ magnetization (e.g. $S^z = \sum_i S^z_i=0$, or $\sum_i \sigma_i = N/2$ in the binary representation) 
required in the ground state of Eq.~\eqref{eq:xy_hamiltonian}.

Although the implementation of symmetries is important for the performance of any quantum algorithm, they have
not yet been implemented in generative models used for quantum state reconstruction.
For the widely-used RBM (see Appendix~\ref{app:rbm}), implementing symmetries is not possible in the block Gibbs sampling algorithm, 
without significant modifications to conventional contrastive divergence (CD$_k$) \cite{hinton_training_2002}.
In contrast, as shown by Hibat-Allah {\it et al}., the autogregessive property of an RNN gives it the 
ability for discrete symmetries to be implemented naturally in the sampling algorithm \cite{Hibat-Allah2020}.
In the variational setting, where the RNN parameters are optimized based on the expectation value 
of the Hamiltonian, the authors of Ref.~\cite{Hibat-Allah2020}
concluded that imposing symmetry improved the accuracy of the ground state thus obtained. 
Here, we ask if a similar performance improvement may be possible in the setting where the RNN is asked to reconstruct
the ground state wavefunction based on measurement data. 

\begin{figure}
    \centering
    \includegraphics[width=\linewidth]{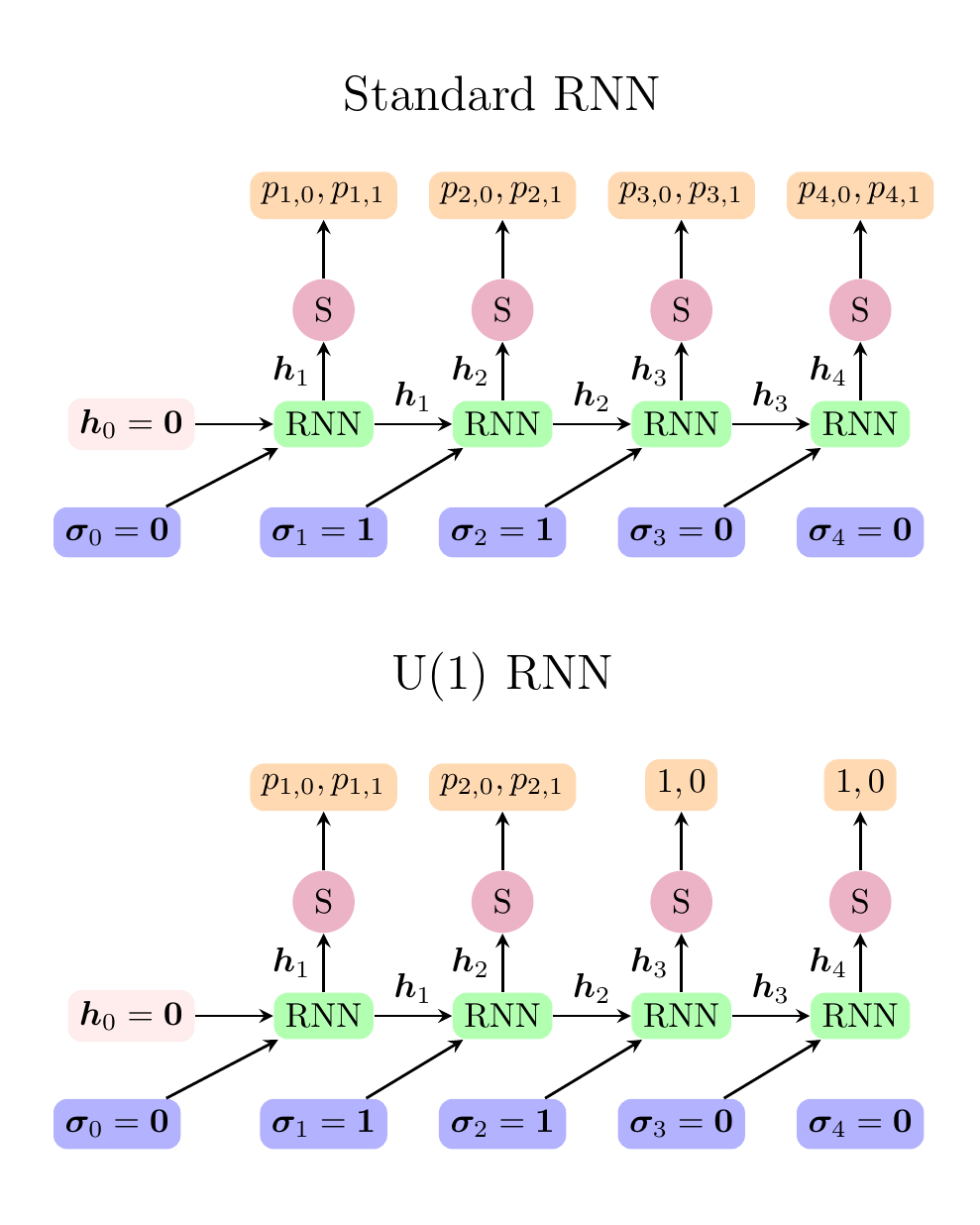}
    \caption{Comparison of a standard RNN and one in which U(1) symmetry is enforced. Here, $p_{i,j}$ refers to the probability $p(\sigma_i = j)$ with $j \in \{0,1\}$. In this example with training sample $\bm{\sigma} = (1,1,0,0)$, the ``standard'' RNN will not impose any restrictions on the output distribution, and therefore might give a non-zero probability of a symmetry-violating configuration. By contrast, the symmetry-enforcing network counts the first two down spins -- the fixed inital input spin $\bm{\sigma}_0 = \bm{0}$ doesn't count -- and then the output is modified to guarantee that the next two spins will be up. Hence $p_{3,0} = p_{4,0} = 1$, and $p_{3,1} = p_{4,1} = 0$. 
    }
    \label{fig:RNN_config}
\end{figure}

Note first, if drawing samples of projective measurements in the $S^z$ basis that conserve zero magnetization, 
there must be exactly $N/2$ spin-up and $N/2$ spin-down states for any even chain length. 
Treating the lattice sites as an arbitrary sequence, this means that
once the cumulative number of either up spins or down spins in the chain reaches $N/2$ we know the remaining spins \textit{must} be of the opposite spin. 
Mathematically, we can describe this following Ref.~\cite{Hibat-Allah2020}. 
At every point in the sequential chain, denote $N_\uparrow$ and $N_\downarrow$ as the cumultive number of up and down spins until that point, respectively. 
Then, for the original network outputs $\bm{y}_i = (p_{i,0}, p_{i,1}) \equiv (p(\sigma_i = 0), p(\sigma_i = 1))$, modify them as follows. Define,
\begin{align} \label{eq:new_probs}
    \widetilde{p}_{i,0} & = p_{i,0} \times \Theta\left(\frac{N}{2} - N_\uparrow\right), \\
    \intertext{and}
    \widetilde{p}_{i,1} & = p_{i,1} \times \Theta\left(\frac{N}{2} - N_\downarrow\right),
\end{align}
where
\begin{equation*} \label{eq:heaviside}
    \Theta(x) = 
    \begin{cases}
        0 & x \leq 0 \\
        1 & x > 0
    \end{cases}
\end{equation*}
is the Heaviside step function. Finally, renormalize to get the modified output
\begin{equation} \label{eq:renorm_probs}
    \widetilde{\bm{y}}_i = \frac{(\widetilde{p}_{i,0}, \widetilde{p}_{i,1})}{\widetilde{p}_{i,0} + \widetilde{p}_{i,1}}.
\end{equation}
This ensures that if there are already $N/2$ up (down) spins, the probability of predicting another up (down) spin is exactly 0. 
We are then guaranteed that the network will predict down (up) spins for all the remaining sites, thereby 
remaining exclusively in the $S^z = 0$ sector.

Training works in exactly the same way with this modification with the exception that the RNN probability distribution $p(\bm{\sigma})$ is replaced by a modified, symmetry-enforcing distribution. 
Also note that for spins on which the symmetry enforcement is applied, these will not contribute to the cost function, and therefore gradient information, since they have probability 1. 
We anticipate that the effect of the projector on the training of the RNN is to alleviate the 
potential vanishing and exploding of the noisy gradients and an ensuing stabilization of the training process. This can be intuitively 
understood from the a sum-of-products structure of the gradient of the loss function ~\cite{10.5555/3042817.3043083}. We first 
note that the projectors effectively shorten the length of the sequences that are used in the calculation of the gradient, which 
eliminates some factors from their calculation that appear multiplicatively~\cite{10.5555/3042817.3043083}.  Due to this
multiplicative structure, the shortening of the sequences effectively stabilizes the training by ammeliorating the 
uncontrolled amplification or supression of the gradients. We note that the effective shortening of the sequences induces an effect 
similar to the improvements brought by the truncated backpropagation (TB) algorithm~\cite{Williams90anefficient}, one of the most most 
practical and successful methods for training RNNs. In TB, a long sequence of $K$ steps is split into $m$ shorter sequences of length $l$ 
(such that $K=m \times l$), each of which is used as a separate training example. While traditional TB leads to the loss of 
potentially important correlations in the data due to the mean-field nature of the algorithm, our approach exactly accounts 
for the correlations due to the magnetization conservation without losing information.  

\section{Results} \label{sec:results}

In this section, we perform numerical experiments designed to test the effectiveness of the symmetry-imposed RNN 
in reconstructing the ground state wavefunction of the spin-1/2 XY model given by Eq.~\eqref{eq:xy_hamiltonian}.
The ground state of the XY model can be efficiently calculated in one dimension using the density matrix renormalization group (DMRG) \cite{ITensor,whiteDensityMatrixAlgorithms1993,whiteDensityMatrixFormulation1992}, and the input data set $\mathcal{D}$ can then be generated via the sampling algorithm of Ferris and Vidal \cite{ferrisPerfectSamplingUnitary2012}. 
For all system sizes, we produce a training dataset of $|\mathcal{D}| = 20000$ independent samples drawn from the DMRG.
Note that we adopt the convention that spin-up refers to $\sigma_i = 0$ and spin-down refers to $\sigma_i = 1$, i.e.~$\sigma_i = 1/2-S^z_i$.
Each training sample is then prepared as a one-hotted vector where $\sigma_i \in \{0,1\}$ is mapped to $\bm{\sigma}_i \in \{(1,0), (0,1)\}$ for each lattice site $i$.

\subsection{Physical observables as training metrics}

An advantage of using RNNs for the reconstruction of quantum states is that the quality of reconstruction can be 
assessed by calculating physical observables (such as the energy) during training, and comparing this to its target 
or ``exact'' value (calculated in this case by DMRG)
\cite{DeVlugt2020,beachQuCumberWavefunctionReconstruction2019, sehayek2019}. The expectation value $\langle \hat{H}\rangle_{\psi_{\text{RNN}}}$ can be 
approximated by an energy estimator $E_{\text{RNN}}$, calculated from a data set $\mathcal{S}$ autoregressively generated from the RNN,
\begin{equation} \label{eq:energy_estimator}
    E_{\text{RNN}} \approx \frac{1}{|\mathcal{S}|} \sum_{\bm{\sigma} \in \mathcal{S}} \sum_{\bm{\sigma}^\prime} \frac{\psi_{\text{RNN}}(\bm{\sigma}^\prime)}{\psi_{\text{RNN}}(\bm{\sigma})} H_{\bm{\sigma}\bm{\sigma}^\prime},
\end{equation}
where $H_{\bm{\sigma}\bm{\sigma}^\prime} = \matrixelement{\bm{\sigma}}{\hat{H}}{\bm{\sigma}^\prime}$. This expression is equivalent to the {\it local observable} formalism used in variational Monte Carlo.

Since all training data sets $\mathcal{D}$ are generated via DMRG,
it is convenient to use the results of these calculations as our target or ``exact'' values for physical observables. 
We choose to monitor the absolute difference in $E_{\text{RNN}}$ and the energy from DMRG (per site),
\begin{equation} \label{eq:energy_difference}
    \varepsilon = \abs{\frac{E_{\text{RNN}} - E_{\text{DMRG}}}{N}},
\end{equation}
for all systems studied in this work. 

Additionally, in our case the fidelity between the DMRG wavefunction (taken to be the target ground state given in Eq.~\eqref{eq:exact_groundstate}) and the RNN wavefunction (Eq.~\eqref{eq:RNN_wavefunction}), can also be monitored during training.
The fidelity is given by,
\begin{equation} \label{eq:fidelity}
    \mathcal{F} = |\braket{\psi_{\text{GS}}}{\psi_{\text{RNN}}}|^2,
\end{equation}
and can be calculated directly using the wavefunction coefficients provided by the DMRG simulation.

\subsection{Numerical results}

\begin{table}[b]
    \caption{Hyperparameters used for training RNNs.}
    \label{tab:rnn_hyperparameters}
    \begin{center}
        \begin{tabular}{ccl}
            \toprule
            Hyperparameter      & Value \\
            \hline
            Hidden units & 100 \\
            Random seed & 1 \\
            Learning rate & 0.001 \\
            Batch size & 50 \\
            \botrule
        \end{tabular}
    \end{center}
\end{table}

Hyperparameters for the RNNs trained in this work can be found in Table~\ref{tab:rnn_hyperparameters}.
As a means of benchmarking performance, the RNN will be compared to the familiar RBM (see Appendix \ref{app:rbm}).
For the RBM, we choose standard CD$_k$ Gibbs sampling given its simplicity, widespread use, and efficiency \cite{pmlr-v48-tosh16, cuevas_novel_2020, carreira-perpinan_contrastive_nodate,sehayek2019,beachQuCumberWavefunctionReconstruction2019,DeVlugt2020,carrasquillaReconstructingQuantumStates2018, Carleo2018NeuralnetworkQS, hintonPracticalGuideTraining2012, salakhutdinovRestrictedBoltzmannMachines2007, torlaigiacomoAugmentingQuantumMechanics2018, torlaiLatentSpacePurification2018, torlaiLearningThermodynamicsBoltzmann2016, torlaiNeuralNetworkQuantumState2018}. 
Hyperparameters used to train the RBMs in this work are presented in Table~\ref{tab:rbm_hyperparameters} of Appendix \ref{app:rbm}.

% \begin{\onecolumn}
    \begin{figure}[t]  % Comparing RNN training
        \centering
        \includegraphics[width=0.85\columnwidth]{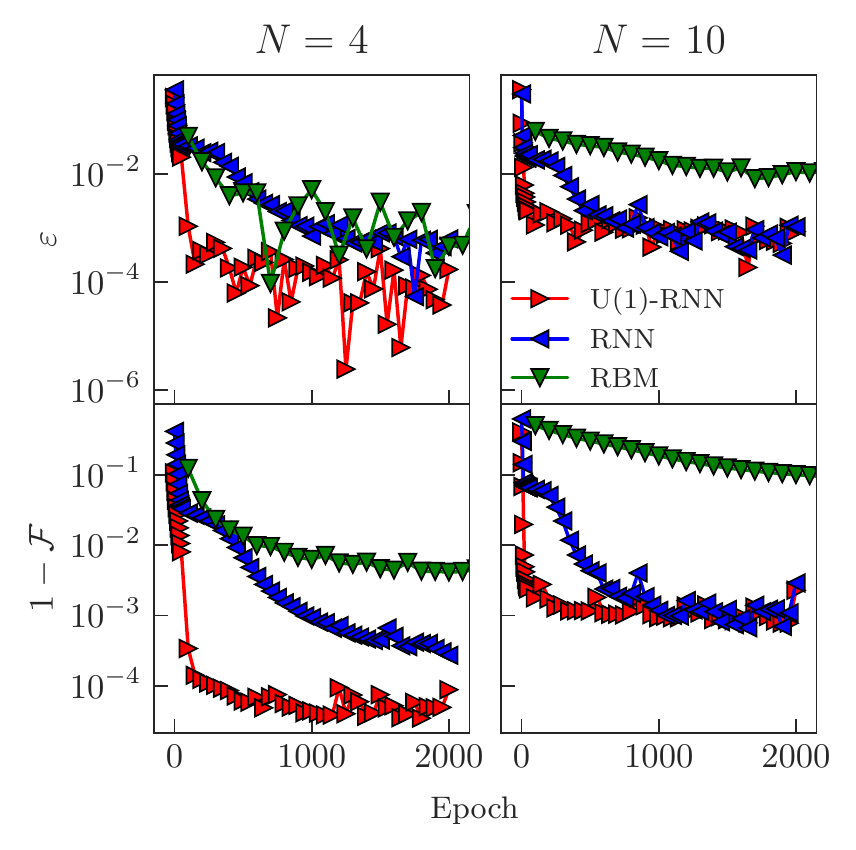}
        \caption{Comparing the energy difference and infidelity during training for $N = 4$ and 10 systems using a recurrent neural network with (U(1)-RNN) and without (RNN) imposing the symmetry, and an RBM. Error bars are omitted for plots of $\varepsilon$ due to the semi-log scale, but are on the order of $10^{-3}$ or smaller.}
        \label{fig:training_compare}
    \end{figure}
% \end{\onecolumn}

\begin{figure}[t]  % Comparing energies vs. epoch at different N
    \centering
    \includegraphics[width=0.9\columnwidth]{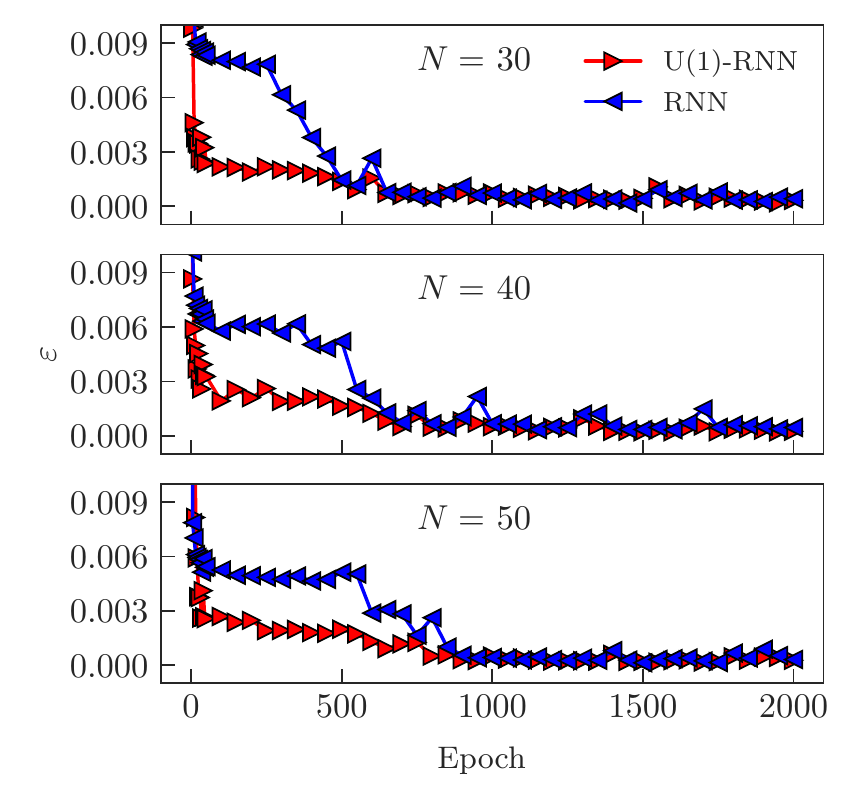}
    \caption{Comparing the energy difference during training for $N = 30$, 40, and 50 systems using a recurrent neural network with (U(1)-RNN) and without (RNN) imposing the symmetry. Errors are on the order of $10^{-3}$ or smaller. 
  }
    \label{fig:energy_compare}
\end{figure}

We begin by studying the accuracy of the RNN reconstruction of the XY model ground state, using as metrics the energy and fidelity.
In the following results, this estimator Eq.~\eqref{eq:energy_estimator} is calculated using
$|\mathcal{S}|=10^4$ projective measurements in the $S^z$ basis, drawn from each generative model (the RBM or RNN). 

In Fig.~\ref{fig:training_compare}, we illustrate the relative energy difference $\varepsilon$, Eq.~\eqref{eq:energy_difference}, and infidelity ($1 - \mathcal{F}$)
for $N = 4$ and $N=10$ using an RBM, a conventional RNN, and an RNN with symmetry enforcement in training (labelled U(1)-RNN). 
Results are plotted as a function of training epoch, where one epoch is defined as a pass over the entire data set $\mathcal{D}$ 
(with a batch size defined in Table~\ref{tab:rnn_hyperparameters}).
In these plots, the RBM clearly does not reach the same reconstruction quality as the RNNs. 
Further, the RBMs were trained out to 20,000 epochs, however we see very little improvement past 2000 epochs. 
Thus, we conclude that the RBMs trained in this work cannot reach a reconstruction quality comparable to the RNNs in reasonable compute time.
Further discussion of this, and results on larger system sizes, is relegated to Appendix \ref{app:rbm}.

The results of Fig.~\ref{fig:training_compare} show that the U(1)-RNN is significantly faster (i.e.~requires fewer epochs) in achieving lower infidelities and $\varepsilon$ 
compared to a conventional RNN without symmetry imposed.
Importantly, in the limit of long training time, we find that reconstruction quality in both RNNs is comparable, 
given the uncertainties involved. 
We explore this for larger system sizes, $N=30$, 40 and 50, in Fig.~\ref{fig:energy_compare}.  Here, it is clear that in each case, 
the U(1)-RNN reaches a small relative energy difference ($\varepsilon < 0.003$) in significantly fewer training epochs than the conventional RNN.
As in the case of smaller $N$, a comparable energy difference is eventually achieved by the conventional RNN, e.g. for 500 or more epochs 
in Fig.~\ref{fig:training_compare}.  After the two models achieve approximately the same energy, further training (e.g.~past 500 epochs) appears to affect 
both RNNs similarly, to within fluctuations.

\begin{figure}  % Comparing energies vs. epoch at different N
    \centering
    \includegraphics[width=0.8\columnwidth]{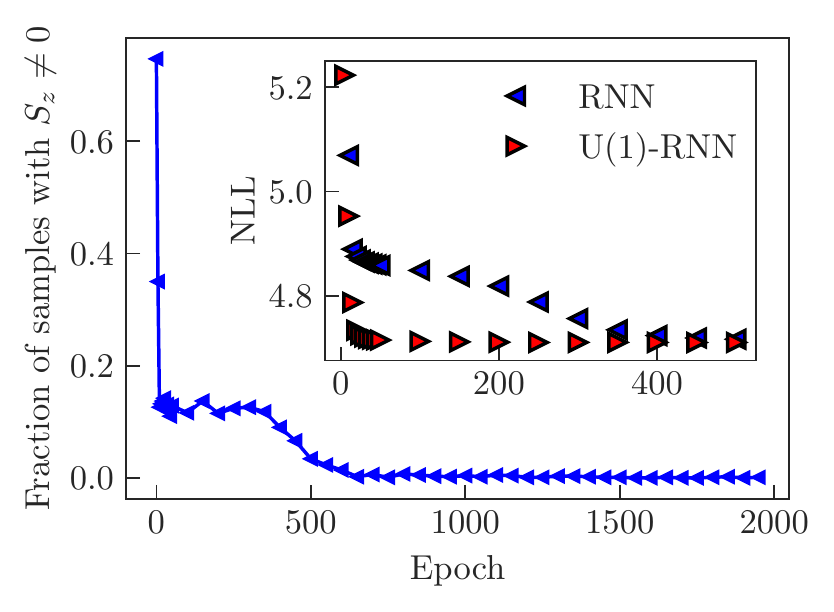}
    \caption{The fraction of samples that a conventional RNN generates outside of the $S_z = 0$ sector for training an $N = 10$ system. The intermediate plateau coincides with a  plateau in fidelity and energy difference (see Fig.~\ref{fig:training_compare}).
    Inset: a comparison of the negative log-likelihood Eq.~\eqref{eq:NLL} between the conventional and U(1)-enforced RNNs.
    }
    \label{fig:bad_sz_samples}
\end{figure}

To further explore the correlation between the implementation of symmetry enforcement, in Fig.~\ref{fig:bad_sz_samples} we illustrate the fraction of 
samples generated by the standard RNN as a function of training, for $N=10$.  Interestingly, after the first few initial epochs of training, the conventional RNN
experiences a plateau where 10\% to 20\% of configurations fall outside of the $S^z=0$ magnetization sector.  This number drops to zero near
epoch 500, which corresponds to the point where the conventional and U(1)-enforced RNNs achieve approximate the same energy and fidelity in
Fig.~\ref{fig:training_compare}. Additionally, as illustrated by the inset of Fig.~\ref{fig:bad_sz_samples}, the plateau of $S^z \neq 0$ samples corresponds in a
higher NLL (Eq.~\eqref{eq:NLL}) for the conventional RNN as compared to the U(1)-enforced version.

As a final illustration of how the symmetry enforcement affects training, we visualize the loss landscape near the optimum for each RNN. 
Building on earlier works \cite{Goodfellow2014, Im2016, Li2018}, we visualize our landscapes by randomly choosing two directions $\bm{\delta}$ and $\bm{\eta}$ 
in the very high-dimensional space of parameters, where each component of the vectors is sampled from a normal distribution, $\mathcal{N}(0,1)$. With these, we plot the function,
\begin{equation}
    f(\alpha, \beta) = L(\bm{\theta^*} + \alpha \bm{\delta} + \beta \bm{\eta}),
\end{equation}
where $\bm{\theta^*}$ are the optimal parameters obtained at the end of training, and $\alpha, \beta$ are scaling parameters 
used to explore the loss function $L$ on the plane defined by $\bm{\theta}^*, \bm{\delta}$, and $\bm{\eta}$. 

\begin{figure}
    \centering
    \includegraphics[width=\linewidth]{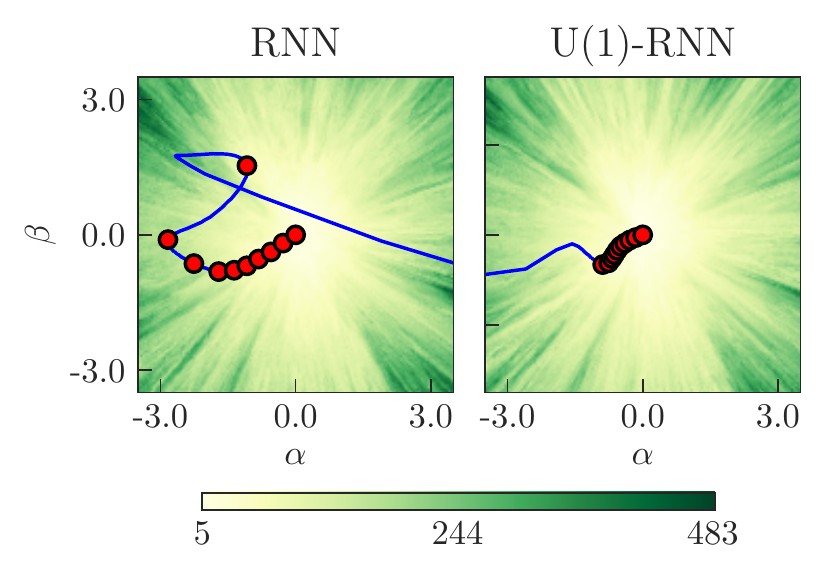}
    \caption{Comparing the loss landscape on a randomly chosen plane in parameter space for the U(1)-RNN and the conventional RNN, training on data from an $N = 10$ system.   The line traces the evolution of the network parameters in training, with the dots evenly spaced every 200 epochs. 
    The first dot illustrated in each path occurs at epoch 200.
    The optimum $\bm{\theta}^*$ found in training sits at the origin. 
    The evolution of infidelity and $\epsilon$ for these parameters appear in Fig.~\ref{fig:training_compare}.
    }
    \label{fig:symm_compare_loss}
\end{figure}

Fig.~\ref{fig:symm_compare_loss} shows the above-defined cross-section of the loss landscape for an RNN with and without symmetry imposed, and the corresponding path that the RNN parameters traverse during training. The system size is $N=10$, with corresponding infidelity and energy differences plotted in Fig.~\ref{fig:training_compare}.  In Fig.~\ref{fig:symm_compare_loss}, the line traces the evolution of the RNN parameters during training, with dots spaced every 200 epochs. 
Although we note that the loss landscapes are different in each case, it is clear that imposing symmetry leads to a more stable training where the RNN parameters approach its respective optimal value in significantly fewer epochs. 
Thus, enforcing the U(1) symmetry causes that optimum to be reached much more quickly than the case where no symmetry is imposed in training.

\section{Discussion} \label{sec:discussion}

In this paper, we have explored the effect of imposing symmetry in a recurrent neural network (RNN) used for quantum state reconstruction.
The system considered is the ground state wavefunction of the spin-1/2 XY model on an open-boundary chain of length $N$.
This model is directly relevant to current trapped ion experiments \cite{Ions1,Ions2,Ion_Review}, which can prepare the XY Hamiltonian on chain lengths of tens of sites or more, and produce a variety of measurements suitable for data-driven state reconstruction.
Due to the U(1) symmetry of the Hamiltonian, the ground state magnetization in the $S^z$ basis is a conserved quantity.  
It is this conservation law that we impose in the autoregressive sampling procedure of the RNN. This is done through a projection operator that constrains the 
RNN output after the point in the sequence for which the symmetry induced constraints fully determine the remaining spin states.

The RNNs in this work are used in the context of state reconstruction from data.
In order to benchmark their performance, we use our knowledge of the exact ground state wavefunction (obtained via DMRG) 
to track the fidelity and relative energy difference as a function of training epoch.
Our main observation is that the imposition of the U(1) symmetry constraint significantly improves the speed at which RNNs 
are able to learn good approximations to the ground state of the XY model. Specifically, such RNNs
require fewer training epochs to reach relatively low infidelities and energy differences compared to a conventional RNN without 
U(1) symmetry imposed.  By examining the tendency of conventional RNNs to produce samples outside of the $S^z = 0$ magnetization sector,
it is clear that the slower training performance in that case correlates to a significant fraction of samples produced outside of this sector.
Upon further training (past approximately 500 epochs in our case), the fidelity and energy metrics of the RNNs with and without symmetry begin to converge,
which corresponds to the point where the conventional RNN begins to produce $S^z = 0$ configurations only.  After this point is reached in training,
both RNNs have similar performance with additional training epochs. We hypothesize that this acceleration arises from the effective reduction of the length of the sequences used in the backpropagation algorithm.  
Since these sequences are constructed multiplicatively, shortening them can reduce any uncontrolled amplification or supression of the gradients.
The resulting stabilization in the gradient evaluation is consistent with our observations of the path traversed by the parameters through the loss landscape,
during the training of the RNN with and without U(1) symmetry.

It will be interesting to extend our study to quantum ground states with negative or complex expansion coefficients, and to  
systems motivated by other NISQ hardware \cite{Preskill2018}.
For example, we anticipate that the exploitation of symmetries 
will also aid the simulation of quantum chemistry on NISQ hardware, since these systems are naturally endowed with a charge conservation symmetry such as the one studied in this work. Symmetry imposition
could be especially beneficial in strategies that use neural-network estimators
to increase the precision of measurements of physical observables in quantum chemistry~\cite{PhysRevResearch.2.022060}. Here, local fermionic observables are transformed into measurements over the entire set of qubits in a quantum computer via the Jordan-Wigner transformation, which increases the statistical errors of important observables significantly.     
Finally, speedups in early epoch training like those identified in this work may be particularly compelling
when combined with NISQ applications that require rapid reconstruction of estimators in an iterative optimization,
such as variational hybrid quantum-classical algorithms \cite{VQEtheory}.

\subsection*{Acknowledgements}

We thank M. Hibat-Allah, R. Islam, and G. Torlai for many enlightening discussions.
Plots in the Results section were generated using the Matplotlib \cite{Hunter:2007} and NumPy \cite{harris_array_2020} packages. 
RGM is supported by the Natural Sciences and Engineering Research Council of Canada (NSERC),
Canada Research Chair (CRC) program, and the Perimeter Institute for Theoretical Physics.
Research at Perimeter Institute is supported in part by the Government of Canada through the Department of Innovation,
Science and Economic Development Canada and by the Province of Ontario through the Ministry of Colleges and Universities.
JC acknowledges support from (NSERC),  the Shared Hierarchical Academic Research Computing Network (SHARCNET), 
Compute Canada, Google Quantum Research Award, and the Canadian Institute for Advanced Research (CIFAR) AI chair program. 
Resources used in preparing this research were provided, in part, by the Province of Ontario, the Government of Canada 
through CIFAR, and companies sponsoring the Vector Institute \url{www.vectorinstitute.ai/#partners}.

\appendix

\section{Restricted Boltzmann Machine (RBM)} \label{app:rbm}

\subsection{Theory} \label{app:rbm_theory}

The RBM is a stochastic neural network comprising two-layers: a {\it visible} layer $\bm{\sigma}$ and a {\it hidden} layer $\bm{h}$. The network contains parameters $\bm{\lambda} = (\bm{W}, \bm{b}, \bm{c})$, where $\bm{W}$ is a weight matrix connecting each node in the visible layer to each node in the hidden layer, and $\bm{b}$ and $\bm{c}$ are external bias fields for the visible and hidden layer, respectively. The RBM defines a probability distribution over its two layers,
\begin{equation} \label{eq:rbm_joint_prob}
    p_{\bm{\lambda}}(\bm{\sigma},\bm{h}) = \frac{e^{-E_{\bm{\lambda}(\bm{\sigma},\bm{h})}}}{Z_{\bm{\lambda}}},
\end{equation}
where
\begin{equation} \label{eq:rbm_joint_energy}
    E_{\bm{\lambda}}(\bm{\sigma},\bm{h}) = -\bm{\sigma}^\intercal \bm{W} \bm{h} - \bm{\sigma}^\intercal \bm{b} - \bm{h}^\intercal \bm{c},
\end{equation}
and normalization
\begin{equation} \label{rbm_partition_fn}
    Z_{\bm{\lambda}} = \sum_{\bm{\sigma},\bm{h}} e^{-E_{\bm{\lambda}(\bm{\sigma},\bm{h})}}.
\end{equation}
The hidden layer is commonly comprised of $n_h$ binary nodes, while in the context of physical systems of $N$ spin-1/2 particles the visible layer corresponds to the $N$ particles.

The marginal distribution over the visible layer is obtained by summing over the hidden layer:
\begin{equation} \label{eq:rbm_marginal_dist}
    p_{\bm{\lambda}}(\bm{\sigma}) = \sum_{\bm{h}} p_{\bm{\lambda}}(\bm{\sigma},\bm{h}) = \frac{e^{-\mathcal{E}_{\bm{\lambda}(\bm{\sigma})}}}{Z_{\bm{\lambda}}},
\end{equation}
where 
\begin{equation} \label{eq:rbm_eff_energy}
    \mathcal{E}_{\bm{\lambda}(\bm{\sigma})} = -\bm{\sigma}^\intercal \bm{b} - \sum_{j=1}^{n_h} \ln(1 + \exp(c_j + \sum_{i=1}^N W_{ij} \sigma_i)).
\end{equation}

An RBM can be trained to reconstruct a target distribution $q({\bm{\sigma}})$ by minimizing the Kullback-Leibler (KL) divergence, 
\begin{equation} \label{eq:KL_divergence}
    \text{KL}_{\bm{\lambda}} = \sum_{\bm{\sigma}} q(\bm{\sigma}) \ln(\frac{q(\bm{\sigma})}{p_{\bm{\lambda}}(\bm{\sigma})}).
\end{equation}
The gradients of the KL divergence with respect to all RBM parameters $\bm{\lambda}$ can be approximated as
\begin{equation} \label{eq:KL_gradient}
    \nabla_{\bm{\lambda}}\text{KL}_{\bm{\lambda}} \approx \expval{\nabla_{\bm{\lambda}}\mathcal{E}_{\bm{\lambda}}(\bm{\sigma})}_{\mathcal{D}} - \expval{\nabla_{\bm{\lambda}}\mathcal{E}_{\bm{\lambda}}(\bm{\sigma})}_{p_{\bm{\lambda}}},
\end{equation}
where $\mathcal{D}$ is a dataset of samples from $q$,
\begin{equation} \label{eq:positive_phase}
    \expval{\nabla_{\bm{\lambda}}\mathcal{E}_{\bm{\lambda}}(\bm{\sigma})}_{\mathcal{D}} = \frac{1}{|\mathcal{D}|} \sum_{\bm{\sigma} \in \mathcal{D}} \nabla_{\bm{\lambda}}\mathcal{E}_{\bm{\lambda}}(\bm{\sigma}), 
\end{equation}
termed the {\it positive phase} of the gradient, and
\begin{equation} \label{eq:negative_phase}
    \expval{\nabla_{\bm{\lambda}}\mathcal{E}_{\bm{\lambda}}(\bm{\sigma})}_{p_{\bm{\lambda}}} \approx \frac{1}{|\Gamma|} \sum_{\bm{\sigma}^{(k)} \in \Gamma} \nabla_{\bm{\lambda}}\mathcal{E}_{\bm{\lambda}}(\bm{\sigma}^{(k)}),
\end{equation}
termed the {\it negative phase} of the gradient. Here, $\Gamma$ denotes samples generated from the RBM via contrastive divergence (CD$_k$) block-Gibbs sampling. Starting from an initial configuration $\bm{\sigma}^{(0)}$ from $\mathcal{D}$, one can sample a hidden layer configuration $\bm{h}^{(0)}$ via
\begin{equation} \label{eq:h_given_v}
    p_{\bm{\lambda}}(h_j = 1 | \bm{\sigma}) = S\left(c_j + \sum_i W_{ij}\sigma_i \right),
\end{equation}
where $S$ is the softmax function. From $\bm{h}^{(0)}$, we may sample a new visible configuration $\bm{\sigma}^{(1)}$ via
\begin{equation} \label{eq:v_given_h}
    p_{\bm{\lambda}}(\sigma_j = 1 | \bm{h}) = S\left(b_i + \sum_j W_{ij}h_j \right).
\end{equation}
Continuing in this alternating fashion $k$ times, we end with a configuration $\bm{\sigma}^{(k)}$. 

The gradients of $\mathcal{E}_{\bm{\lambda}}$ with respect to all RBM parameters are as follows.
\begin{subequations} \label{eq:gradients}
    \begin{align}
        \frac{\partial \mathcal{E}_{\bm{\lambda}}\left(\bm{\sigma}\right)}{\partial W_{ij}} & = - p_{\bm{\lambda}}\left(h_j = 1 \vert \bm{\sigma} \right) \sigma_i , \\
        \frac{\partial \mathcal{E}_{\bm{\lambda}}\left(\bm{\sigma}\right)}{\partial c_j}     & = - p_{\bm{\lambda}}\left(h_j = 1 \vert \bm{\sigma} \right),              \\
        \intertext{and}
        \frac{\partial \mathcal{E}_{\bm{\lambda}}\left(\bm{\sigma}\right)}{\partial b_{i}}  & = -\sigma_{i}.
    \end{align}
\end{subequations}

\subsection{RBMs for Quantum State Reconstruction}

Given a target quantum state we wish to reconstruct, an RBM as outlined in Sec.~\ref{app:rbm_theory} may be employed to reconstruct it if the target state has no sign structure and can therefore be written as
\begin{equation} \label{eq:rbm_wvfn}
    \psi(\bm{\sigma}) = \sqrt{q(\bm{\sigma})}.
\end{equation}
Given Eq.~\eqref{eq:rbm_marginal_dist}, the RBM defines a trial state
\begin{equation}
    \psi_{\bm{\lambda}}(\bm{\sigma}) = \sqrt{p_{\bm{\lambda}}(\bm{\sigma})}.
\end{equation}
Therefore, in reconstructing $q(\bm{\sigma})$ we also reconstruct the target state $\psi(\bm{\sigma})$.

All RBMs trained in this work were written using the QuCumber package \cite{beachQuCumberWavefunctionReconstruction2019} and employed SGD to perform the optimization of the hyperparameters in Tab.~\ref{tab:rbm_hyperparameters}. Note that one gradient update is comprised over many smaller updates over mini-batches of $\mathcal{D}$. The size of the mini-batches used to calculate Eq.~\eqref{eq:positive_phase} is called the {\it positive batch size}, while $|\Gamma|$ in Eq.~\eqref{eq:negative_phase} is called the {\it negative batch size}.

Six different random seeds to initialize the RBM parameters were also investigated. For all RBM results in this work, we chose the random seed that gave the best results in terms of the training metric (Eq.~\eqref{eq:energy_difference} and Eq.~\eqref{eq:fidelity}). Regarding the hidden layer size $n_h$ in the RBMs, it was observed in preliminary calculations that hidden layer sizes in Tab.~\ref{tab:rbm_hyperparameters} gave the best numerically stable results.

\begin{table}
    \caption{Hyperparameters used for training RBMs.}
    \label{tab:rbm_hyperparameters}
    \begin{center}
        \begin{tabular}{ccl}
            \toprule
            Hyperparameter      & Value \\
            \hline
            $n_h$ & 10 [$N = 2$], 50 [$N = 4$], 100 [$N > 4$] \\
            Random seeds & 7777 [$N = 2$], 9999 [$N = 4, 30, 50$], \\ & 2222 [$N = 6$], 1234 [$N = 8, 10, 20$], \\ & 1357 [$N = 16, 40$] \\
            Learning rate              & $0.01 \times 0.999^t$, where $t$ is epoch number \\
            Positive batch size  & 100 \\
            Negative batch size & 200 \\
            $k$ Gibbs steps for $\Gamma$ & 100 \\
            \botrule
        \end{tabular}
    \end{center}
\end{table}

\begin{figure}  % Comparing energies vs. N
    \centering
    \includegraphics[width= 0.8\columnwidth]{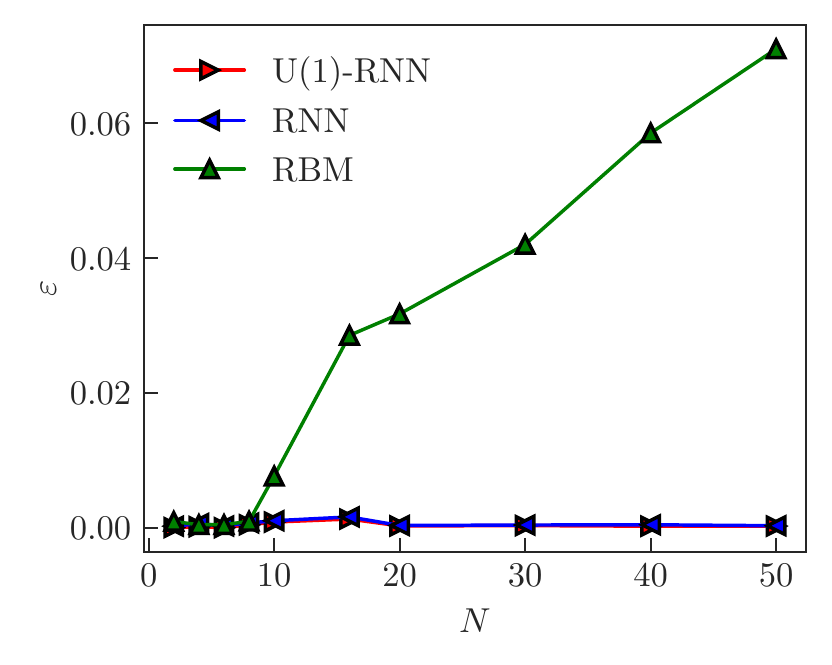}
    \caption{Comparing the energy difference for larger generative models.
    After 2000 epochs of training were completed, $10^4$ samples were generated from the U(1)-RNN, RNN, and RBM to respectively calculate $\varepsilon$. 
    All networks were sufficiently converged at this time and error bars are smaller than the plot markers.
    After this time in the epoch regime, the RNN and U(1)-RNN yield nearly identical values of $\varepsilon$.
    In contrast, the RBM's $\varepsilon$ is drastically higher than both of the RNNs for chain lengths greater than 8.
    }
    \label{fig:energy_vs_N}
\end{figure}

In Figure \ref{fig:energy_vs_N}, we show a plot of the energy difference across several different system sizes after 2000 epochs of training, i.e. where the optimal parameters $\bm{\theta^*}$ have been reached, for the U(1) - RNN, RNN, and RBM.
For RNNs, the energy difference at this time shows minimal improvement regardless of whether symmetry was imposed or not.
What is interesting here is the drastically different behaviour of the RBM.
For RBMs in this work, it was observed that hidden layer sizes greater than 100 yielded worse results (i.e. larger $\varepsilon$ and infidelity) than what is reported.
Not only this, but with larger hidden layer sizes often came less stable numerics; the training metrics did not come to a stable equilibrium.
To try and ameliorate the training metrics and stabilize the numerics when larger hidden layer sizes were used, a quicker learning rate decay was employed in order to weaken the gradient updates. 
Unfortunately, this also did not bear any fruit.

Although exact properties of the ground state of the XY model can be extracted from RBMs analytically \cite{rrapaj_exact_2020, deng_quantum_2017}, the observation of the difficulty in training RBMs with widely-used and robust methods for reconstructing the ground state of the XY model shows that the optimization problem is non-trivial for larger system sizes.
A similar result was reported in Ref.~\cite{DeVlugt2020}.
With relatively little pre-training effort, we were able to acheive much better training metrics and numerical stability with RNNs.
To summarize, RNNs seem to be much easier to optimize than RBMs in this context.

\section{Gated Recurrent Networks} \label{appendix:GRUs}

Due to the nature of their architecture, basic RNNs have trouble learning long-distance dependencies between qubits since the gradients with respect to RNN parameters tend to either explode or vanish. To overcome this, we employ the gated recurrent unit (GRU) introduced in Ref.~\cite{cho-etal-2014-properties}. These process data in a similar way to RNNs, defining:
\begin{align*}
    \bm{z}_i & = \sigma(W_z \bm{\sigma}_{i-1} + U_z \bm{h}_{i-1} + \bm{b}_z) \\
    \bm{r}_i & = \sigma(W_r \bm{\sigma}_{i-1} + U_r \bm{h}_{i-1} + \bm{b}_r) \\
    \hat{\bm{h}}_i & = \tanh (W_h \bm{\sigma}_{i-1} + U_h (\bm{r}_i \odot \bm{h}_{i-1}) + \bm{b}_h) \\
    \bm{h}_i & = (1 - \bm{z}_i) \odot \bm{h}_{i-1} + \bm{z}_i \odot \hat{\bm{h}}_i,
\end{align*}
where $\sigma$ refers to the sigmoid function, and $\odot$ denotes element-wise multiplication. The new hidden state is an interpolation between the previous hidden state $\bm{h}_{i-1}$ and a proposed hidden state $\hat{\bm{h}}_i$, weighted by the update gate $\bm{z}_i$. The reset gate $\bm{r}_i$ controls to what extent the proposed hidden state ``remembers'' the previous hidden state. So if $\bm{r}_i = 0$, the unit acts as though this is the first element of the sequence, making the proposed state ``forget'' the part of the sequence already encoded in $\bm{h}_{i-1}$. The output layer remains unchanged from the basic RNN.

The matrices $W_{h,r,z}$, and $U_{h,r,z}$ and biases $\bm{b}_{h,r,z}$ are all filled with trainable parameters. For hidden units with dimension $n_h$ and inputs with dimension $n_v$, then $W_{z,r,h} \in \mathbb{R}^{n_h \times n_v}$, $U_{z,r,h} \in \mathbb{R}^{n_h \times n_h}$ and $\bm{b}_{z,r,h} \in \mathbb{R}^{n_h}$. The rest of the training procedure remains unchaged from how it operators in a regular RNN. 

\bibliographystyle{apsrev4-1}
\bibliography{bibliography}

\end{document}